\documentclass[a4paper,11pt]{article}
\pdfoutput=1 

\usepackage{jinstpub} 

\usepackage{lineno}
\usepackage{siunitx}

\title{\boldmath Precise timing and recent advancements with segmented anode PICOSEC Micromegas prototypes }

\collaboration{%
\includegraphics[height=17mm]{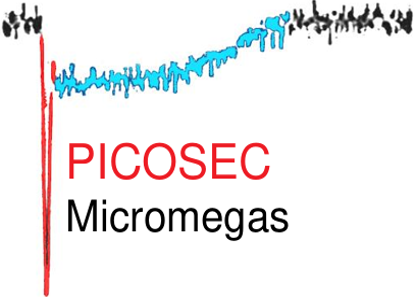}\\[6pt]
PICOSEC-Micromegas collaboration}

\author[d,e,1,2]{I. Manthos,\note{Corresponding author}\note{Now at University of Birmingham}}
\author[a]{S. Aune,}
\author[b,3]{J. Bortfeldt,\note{Now at LMU, Munich}}
\author[b]{F. Brunbauer,}
\author[b]{C. David,}
\author[a]{D. Desforge,}
\author[f]{\mbox{G. Fanourakis,}}
\author[h]{M. Gallinaro,}
\author[l]{F. Garc\'{i}a,}
\author[a]{I. Giomataris,}
\author[j]{T. Gustavsson,}
\author[a]{F.J. Iguaz,}
\author[d,e]{\mbox{A. Kallitsopoulou,}}
\author[a]{M. Kebbiri,}
\author[d,e]{K. Kordas,}
\author[d,e]{C. Lampoudis,}
\author[a]{P. Legou,}
\author[b]{M. Lisowska,}
\author[c]{J. Liu,}
\author[b,3]{M. Lupberger,\note{Now at University of Bonn}}
\author[a]{O. Maillard,}
\author[d,e]{{I. Maniatis,}}
\author[b]{H. M\"{u}ller,}
\author[b]{E. Oliveri,}
\author[a]{\mbox{T. Papaevangelou,}}
\author[d]{K. Paraschou,}
\author[k]{M. Pomorski,}
\author[c]{B. Qi,}
\author[b]{F. Resnati,}
\author[b]{L. Ropelewski,}
\author[d,e]{\mbox{D. Sampsonidis,}}
\author[b]{L. Scharenberg,}
\author[b]{T. Schneider,}
\author[a]{L. Sohl,}
\author[b]{M. van Stenis,}
\author[d,e]{A. Tsiamis,}
\author[g]{Y. Tsipolitis,}
\author[d,e]{S.E. Tzamarias,}
\author[b]{A. Utrobicic,}
\author[i,5]{R. Veenhof,\note{Also at MEPhI \& Uludag University}}
\author[c]{X. Wang,}
\author[b,6]{S. White,\note{Also at University of Virginia}}
\author[c]{\mbox{Z. Zhang}}
\author[c]{and Y. Zhou}

\affiliation[a]{IRFU, CEA, Universit´e Paris-Saclay, F-91191 Gif-sur-Yvette, France}
\affiliation[b]{European Organization for Nuclear Research (CERN), CH-1211 Geneve 23, Switzerland}
\affiliation[c]{State Key Laboratory of Particle Detection and Electronics, University of Science and Technology of China, Hefei CN-230026, China}
\affiliation[d]{Department of Physics, Aristotle University of Thessaloniki, University Campus, GR-54124, Thessaloniki, Greece.}
\affiliation[e]{Center for Interdisciplinary Research and Innovation (CIRI-AUTH), Thessaloniki 57001, Greece.}
\affiliation[f]{Institute of Nuclear and Particle Physics, NCSR Demokritos, GR-15341 Agia Paraskevi, Attiki, Greece}
\affiliation[g]{National Technical University of Athens, Athens, Greece}
\affiliation[h]{Laborat\'{o}rio de Instrumentac\~{a}o e F\'{i}sica Experimental de Part\'{i}culas, Lisbon, Portugal}
\affiliation[i]{RD51 collaboration, European Organization for Nuclear Research (CERN), CH-1211 Geneve 23, Switzerland}
\affiliation[j]{LIDYL, CEA, CNRS, Universit Paris-Saclay, F-91191 Gif-sur-Yvette, France}
\affiliation[k]{CEA-LIST, Diamond Sensors Laboratory, CEA Saclay, F-91191 Gif-sur-Yvette, France}
\affiliation[l]{Helsinki Institute of Physics, University of Helsinki, FI-00014 Helsinki, Finland}

\emailAdd{ioannis.manthos@cern.ch}

\abstract{Timing information in current and future accelerator facilities is important for resolving objects (particle tracks, showers, etc.)  in extreme large particles multiplicities on the detection systems. The PICOSEC Micromegas detector has demonstrated the ability to time 150\,GeV muons with a sub-25\,ps precision. Driven by detailed simulation studies and a phenomenological model which describes stochastically the dynamics of the signal formation, new PICOSEC designs were developed that significantly improve the timing performance of the detector. PICOSEC prototypes with reduced drift gap size ($\sim$\SI{119}{\micro\metre}) achieved a resolution of 45\,ps in timing single photons in laser beam tests (in comparison to 76\,ps of the standard PICOSEC detector). Towards large area detectors, multi-pad PICOSEC prototypes with segmented anodes has been developed and studied. Extensive tests in particle beams revealed that the multi-pad PICOSEC technology provides also very precise  timing, even when the induced signal is shared among several neighbouring pads. Furthermore, new signal processing algorithms have been developed, which can be applied during data acquisition and provide real time, precise timing.}


\proceeding{12$^{\text{th}}$ International Conference on Position Sensitive Detectors \\
 12 - 17 September 2021\\
University of Birmingham, UK}

\begin{document}
\maketitle
\flushbottom

\section{Introduction}
\label{sec:intro}
Very high beam luminosity in future particle physics experiments will generate an enormous particle flux in the detectors. Detectors providing timing resolution of tens of picoseconds for Minimum Ionising Particles (MIP) will be required to provide vertex separation in beam collisions with many overlapping interactions  (pile-up). Furthermore, such detectors should be robust against the extreme high particle multiplicities generated in these interactions. Micromegas detectors have demonstrated such a robustness, whilst  PICOSEC-Micromegas has already demonstrated 24\,ps timing resolution \citep{a}. PICOSEC is a Micromegas detector, with a Cherenkov radiator and a photocathode coupled on top of the gaseous volume, while the drift region length is as small as the amplification region (\SI{200}{\micro\metre}).  Due to the extremely small width of the detector the probability that the charged particle produces an ionisation is negligible. Instead, synchronous primary electrons produced by the Cherenkov photons on the photocathode enter the detector volume and generate the signal, thus reducing the time jitter of the first interaction from several ns to a few ps. Also, due to the increased electric drift field, amplification also occurs in the drift gap (pre-amplification avalanche). \\
The segmented anode multi-pad PICOSEC-Micromegas is the approach towards large area detector. Detailed studies performed in muon test beam demonstrated that such a device provides excellent time resolution and detection efficiency \citep{b}. Results on the single and multi-pad prototypes are presented in Section \ref{sec:prototypes}. \\
A phenomenological model, based on GARFIELD++ simulations, has been developed to describe the signal formation mechanism \citep{c}. This model drove the design of a new PICOSEC-Micromegas prototype, with shorter drift region length at \SI{119}{\micro\metre} and higher drift fields, which has performed with a resolution of 44\,ps in timing single photoelectrons (p.e.) \citep{d}. The same detector performed with  $<$6\,ps resolution in timing laser pulses producing $ \sim$70 photoelectrons on the photocathode, as described in Section \ref{sec:newres}. Section \ref{alttiming} discusses the application of alternative signal processing methods, aiming to retain the very high timing resolution of the detector while minimising the necessary experimental information to be stored, as well as providing the possibility of on-line timing.
\section{Results from single and multipad PICOSEC Micromegas prototypes}
\label{sec:prototypes}
\begin{figure}[h] 
\centering
\begin{minipage}{.45\textwidth}
 \includegraphics[width=.88\linewidth]{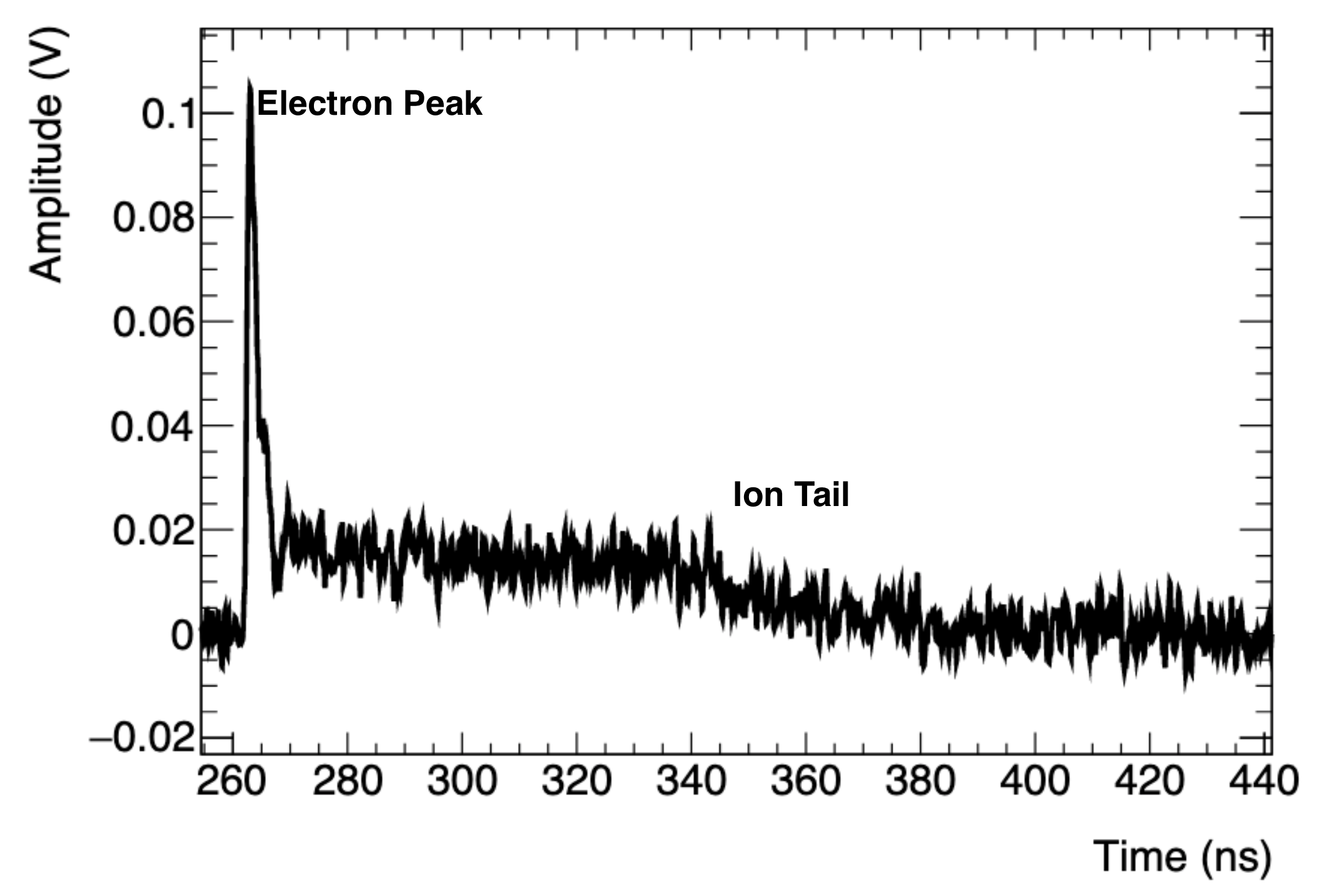}
\end{minipage}
\centering
\begin{minipage}{.4\textwidth}
\includegraphics[width=.8\linewidth]{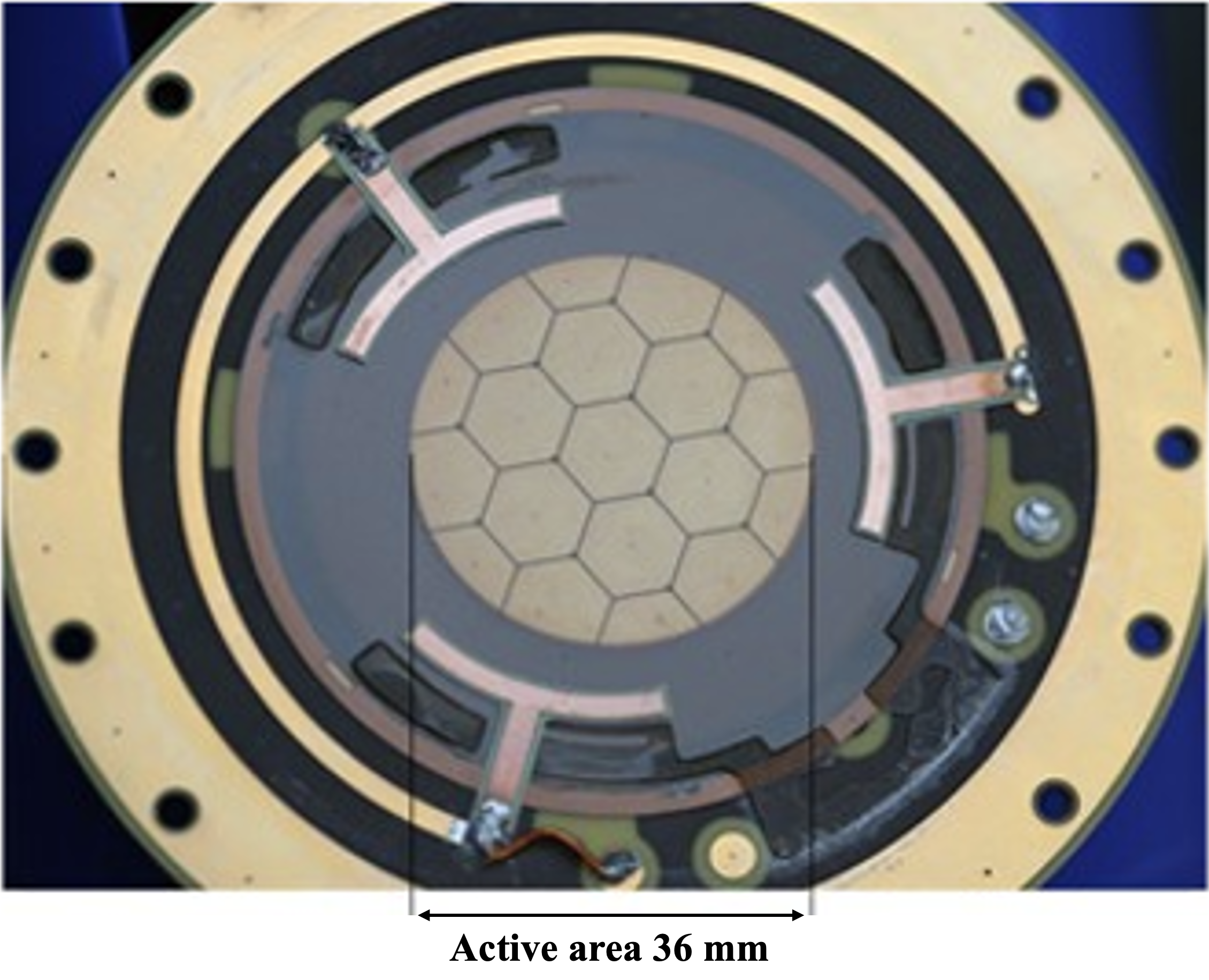}
\end{minipage}
\caption{(Left) A digitised (inverted) PICOSEC waveform, where the Electron Peak and the Ion Tail are indicated. (Right) The multi-pad PICOSEC prototype. \label{fig:fig1}}
\end{figure}
The PICOSEC Micromegas signal consists of a sharp Electron Peak (e-Peak) produced by the fast moving electrons, followed by the Ion Tail produced by the slower moving ions, as presented in Figure \ref{fig:fig1} (left). The timing performance of the single-channel PICOSEC Micromegas prototype  was evaluated using an attenuated pulsed laser beam at the IRAMIS facility at CEA-Saclay and at the SPS H4 beam line at CERN with 150\,GeV muon test beam \citep{a}. In the laser tests, a photodiode with 13\,ps precision was used as time reference and a timing resolution of 76.0$\pm$0.4\,ps for single p.e. was achieved, at +450\,V/-425\,V anode/drift operating voltages. In muon tests, a resolution of 24.0$\pm$0.3 ps was measured for timing the arrival of incoming muons, when the PICOSEC was operated at +275\,V/-475\,V anode/drift voltages, using a MCP detector with 5\,ps timing resolution as reference. The PICOSEC was filled with COMPASS gas mixture (Ne (80\%) + C$_{2}$H$_{6}$ (10\%) + CF$_{4}$ (10\%)), whilst the signal passed through a CIVIDEC pre-amplifier and was digitized by a high-bandwidth, fast digital oscilloscope (20\,GS/s). The digitised waveforms were analysed offline and the leading edge of the e-Peak was fit to a logistic function. The Constant Fraction Discrimination (CFD) technique was applied to the fitted leading edge at the 20\% of the e-Peak maximum. The Signal Arrival Time (SAT) was defined relative to the time-reference device, while the PICOSEC timing resolution was defined as the RMS of the SAT distribution. \\
Detailed studies have also been performed with a multi-pad PICOSEC prototype (Figure \ref{fig:fig1}, right) \citep{b}, using data collected at the SPS H4 beam line at CERN, and performing a similar offline analysis as described above. A timing resolution of 25.8$\pm$0.6\,ps was achieved when the muons impact the detector close to anode pad centres, while the timing resolution deteriorates to 32.2$\pm$0.5\,ps when the muon track passes close to the pad corners and the Cherenkov ring is shared among three pads.  
\section{Performance of PICOSEC prototypes with reduced drift gap }
\label{sec:newres}
\begin{figure}[h] 
\centering
\begin{minipage}{.44\textwidth}
 \includegraphics[width=0.85\linewidth]{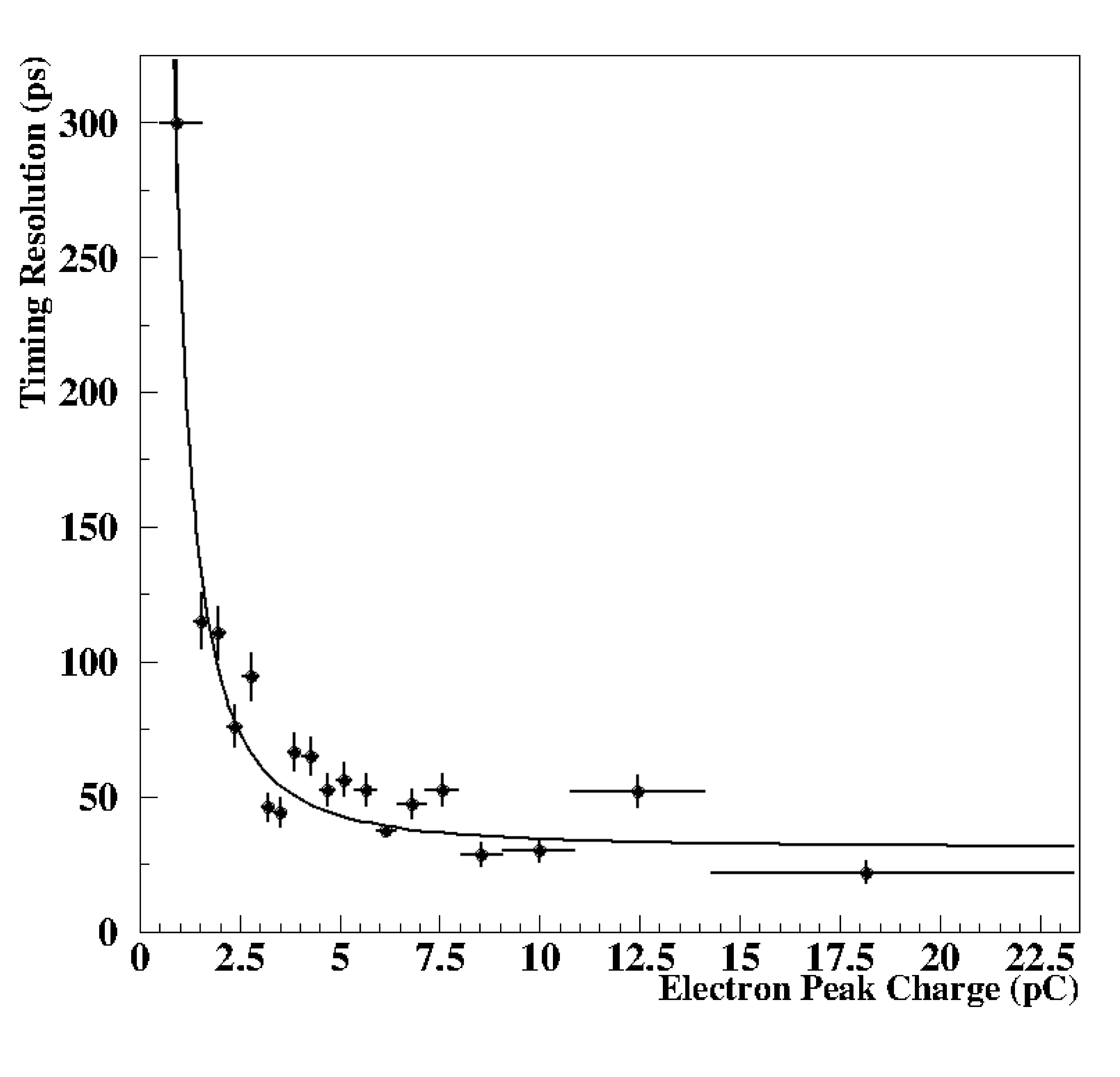}
\end{minipage}
\centering
\begin{minipage}{.44\textwidth}
\includegraphics[width=0.85\linewidth]{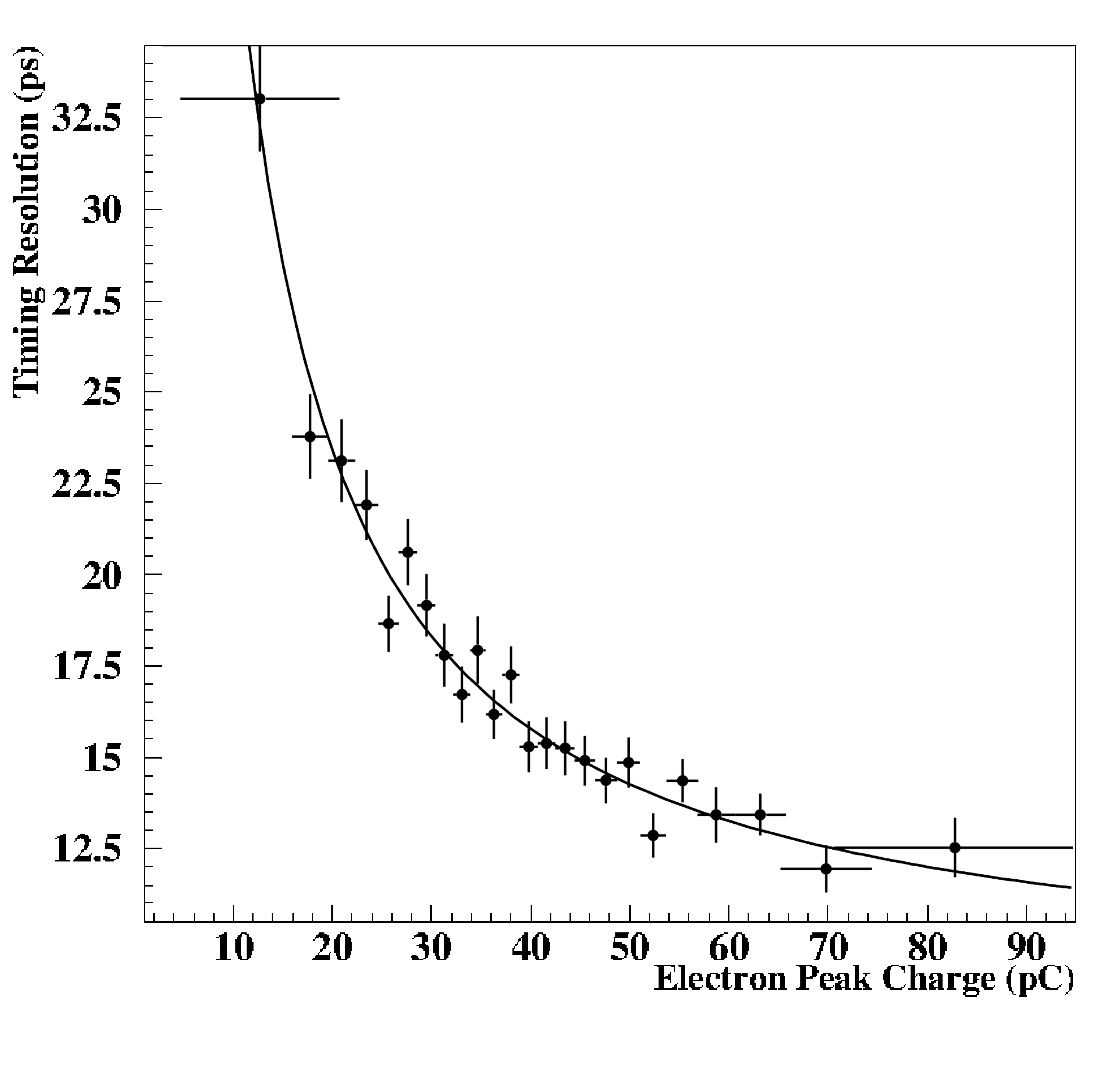}
\end{minipage}
\caption{The timing resolution as a function of the e-Peak charge (Left) on the response of the PICOSEC detector to single p.e. and (Right) on the response to 7.8 p.e. In both cases a power law fit is performed.\label{fig:fig2}}
\end{figure}
\begin{figure}[h] 
\centering
\begin{minipage}{.44\textwidth}
 \includegraphics[width=0.85\linewidth]{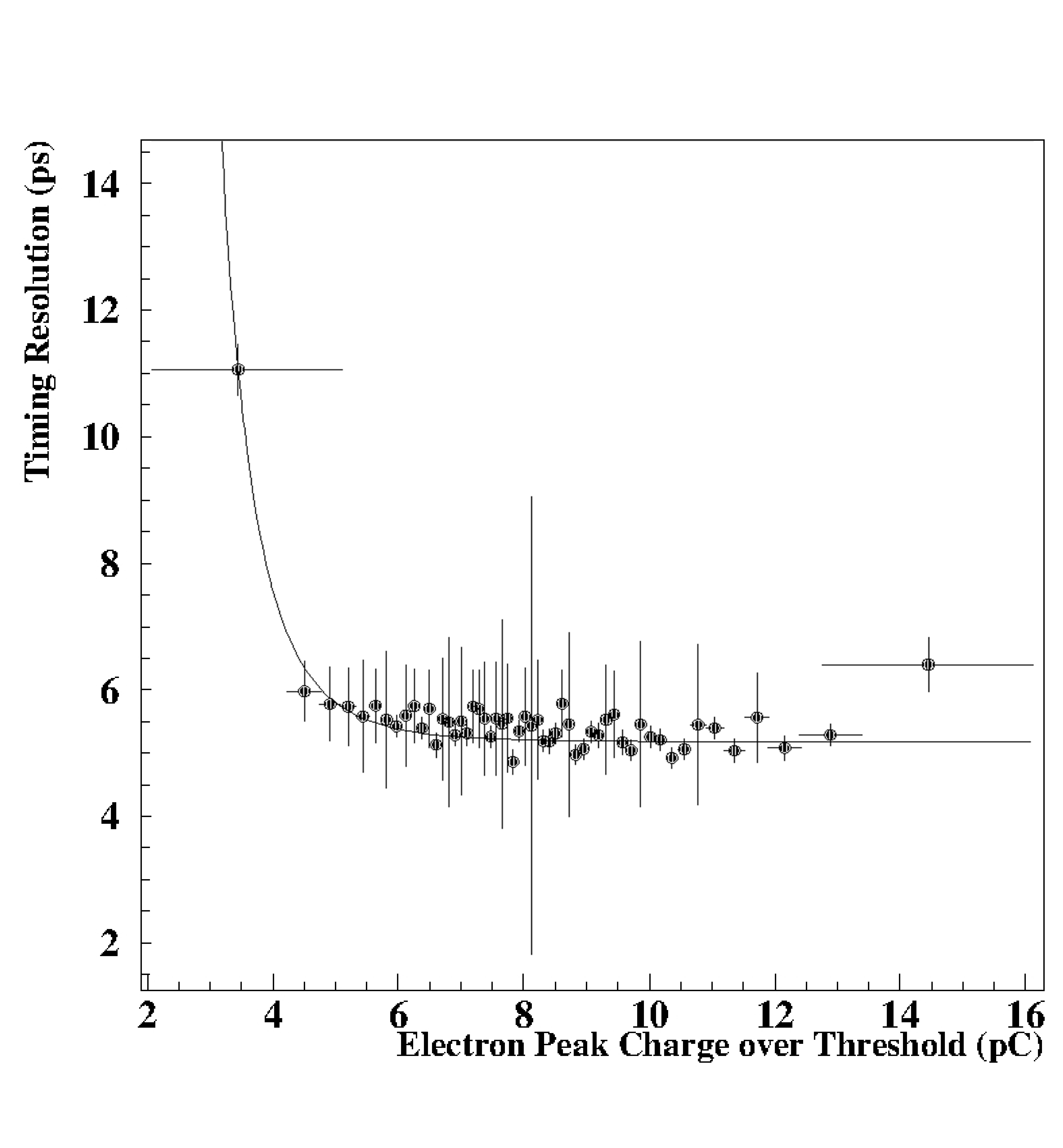}
\end{minipage}
\centering
\begin{minipage}{.44\textwidth}
\includegraphics[width=0.85\linewidth]{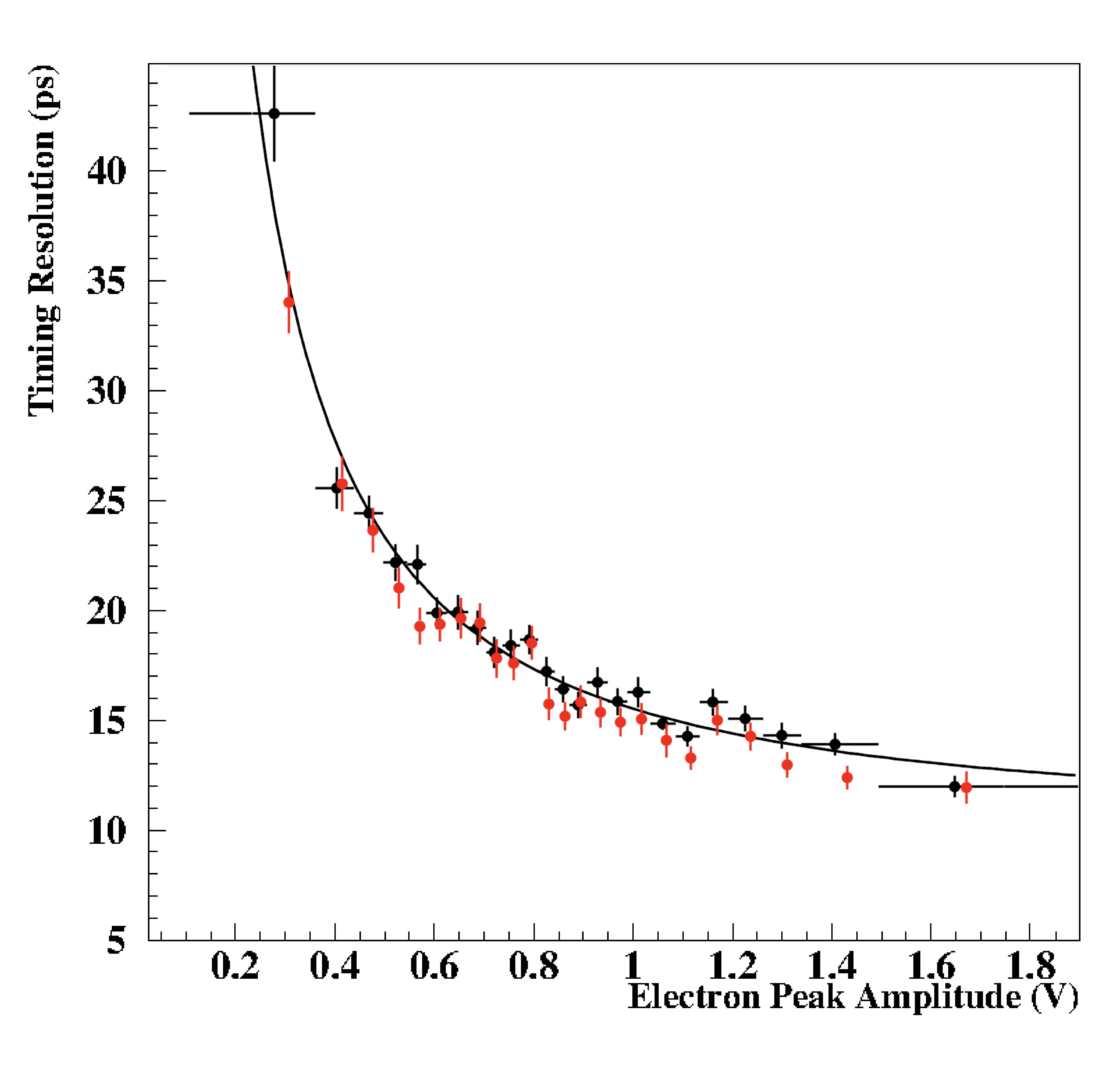}
\end{minipage}
\caption{(Left) Timing resolution as a function of the e-Peak charge when detector responds to a mean value of $\sim$70 p.e. (Right) Resolution as a function of e-Peak amplitude after time-walk correction, using CFD (red) and multi Charge over Threshold (black) timing techniques. The line is a power law fit of the black points.  \label{fig:fig3}}
\end{figure}

The PICOSEC performance is successfully reproduced in detailed simulation studies \citep{c}, based on GARFIELD++ package that included the electronic response of the detector and the noise contribution. The agreement between simulation and experimental data was further exploited in order to identify the microscopic physical variables that determine the observed timing characteristics. To acquire a deeper comprehension on the mechanisms that determine the signal formation, a phenomenological model was developed \cite{c}. The model describes the propagation of the pre-amplification avalanche, showing that the length and multiplication of the avalanche in the drift region is the dominant factor in the timing performance. It also reproduces the PICOSEC timing characteristics equally well as the detailed GARFIELD++ simulations. Driven by the phenomenological model, modified PICOSEC prototypes were designed and tested in laser beam at the IRAMIS facility at CEA-Saclay \cite{d}. Using the aforementioned experimental setup, the PICOSEC was equipped with an Aluminium photocathode and the timing performance was studied with a fast photodiode (<3\,ps) to serve as the time reference. Optimum results achieved with a $\sim$\SI{119}{\micro\metre} drift gap PICOSEC (anode region remained to $\sim$\SI{128}{\micro\metre}), operating at +275\,V/-525\,V anode/drift voltage settings. As presented in Figure \ref{fig:fig2} (left), a 44$\pm$1\,ps timing resolution achieved on the response of the PICOSEC detector to single p.e. Operating with higher laser intensities, the photocathode yield was estimated using a consistent and unbiased method \cite{e}. The offline data analysis resulted in a timing resolution of 18.3$\pm$0.2\,ps when the detector responds to 7.8$\pm$0.1\,p.e. (Figure \ref{fig:fig2}, right). The performance of the PICOSEC detector to an increased intensity  laser beam entering the detector volume (mean value of $\sim$70 p.e. )  is better than 6\,ps,  as shown in Figure \ref{fig:fig3} (left).
\section{Alternative timing techniques for scalable electronics}
\label{alttiming}
In order to minimise the necessary experimental information to be stored, alternative signal processing methods have been developed. The first method is based on Constant Threshold Discrimination timing and time-walk corrections by utilising fast integration of the e-Peak waveforms above certain thresholds. The other timing technique is using ANN trained by simulated waveforms generated as described in \citep{f,g}. The potential of the developed timing techniques has been evaluated by using the data set where the PICOSEC detector responds to light pulses generating an average of 7.8\,p.e.\\
Both methods result to a global timing accuracy  of 18.3$\pm$0.2\,ps (e.g.  Figure \ref{fig:fig3}, right, shows the timing resolution as a function of the e-Peak amplitude), which equals the accuracy offered by the full, offline signal processing analysis.
\section{Conclusions}
PICOSEC-Micromegas detector has demonstrated an excellent timing resolution of 24.0$\pm$0.3\,ps for timing the arrival of a MIP. Driven by a phenomenological model, reduced drift gap prototype demonstrates the timing of laser pulses arrival that generates $\sim$7.8 p.e. per pulse on the detector, with 18.3$\pm$0.2\,ps accuracy. Furthermore, aiming to minimise the  necessary information to be stored  alternative timing techniques were developed and evaluated. The timing resolution is retained, proving the feasibility of scalable electronics for large scale detectors.
\acknowledgments
We acknowledge the support of the RD51 collaboration, in the framework of RD51 common projects. We kindly thank the IRAMIS facility at CEA-Saclay for providing the laser test beam.

\end{document}